\documentclass[a4paper,11pt]{article}

\usepackage{jheppub}
\usepackage{graphics}

\def\mll{$m_{ll}^{\text{edge}}$}

\def\mllqedge{$m_{llq}^{\text{edge}}$}
\def\mllqthr{$m_{llq}^{\text{thr}}$}
\def\mlqlow{$m_{lq(\text{low})}$}
\def\mlqhigh{$m_{lq(\text{high})}$}
\def\mt2{$M_{T2}$}

\newlength{\wth}
\newcommand{\onegraphwcntrs}[2]{\unitlength=1.1in
\begin{picture}(2.25,2.25)
 \put(-0.5,2.25){\epsfig{file=#1,width=0.65\wth,angle=270}}
 \put(-0.5,2.25){\epsfig{file=#2, width=0.65\wth,angle=270}}
\end{picture}}

\newcommand{\twographs}[2]{\unitlength=1.1in
\begin{picture}(5.4,2.25) %(0,0.25)
  \put(-0.5,2.25){\epsfig{file=#1, width=0.65 \wth, angle=270}}
    \put(2.3,2.25){\epsfig{file=#2, width=0.685 \wth, angle=270}}
    \put(0.0,2.0){(a)}
    \put(2.8,2.0){(b)}
\end{picture}}

\title{Supersymmetry With Prejudice: Fitting the Wrong Model to LHC Data}
\author[a]{B C Allanach,}
\affiliation[a]{DAMTP, CMS, University of Cambridge, Wilberforce Road, Cambridge CB3 0WA, United Kingdom}
\emailAdd{B.C.Allanach@damtp.cam.ac.uk}

\author[b]{Matthew J Dolan}
\affiliation[b]{IPPP, 
Department of Physics,
University of Durham,
Science Laboratories,
South Rd,
Durham DH1 3LE, United Kingdom}
\emailAdd{m.j.dolan@durham.ac.uk}

\abstract{We critically examine interpretations of hypothetical
  supersymmetric LHC signals, fitting to alternative wrong models of
  supersymmetry breaking. The signals we consider are some of the most
  constraining on the sparticle spectrum: invariant mass distributions with
  edges and   end-points from the golden cascade decay chain $\tilde{q}_L \to
 q \chi_2^0  \left(   \to \tilde{l}^{\pm} l^{\mp} q \right) \to \chi_1^0 l^+
  l^- q$. We assume a CMSSM point to be the `correct' one, but fit the signals
  instead 
  with minimal gauge mediated supersymmetry breaking models (mGMSB) with a
  neutralino 
  quasi-stable lightest supersymmetric particle, minimal anomaly mediation
  (mAMSB) and
  large volume string compactification models (LVS). 
  mAMSB and LVS can be unambiguously discriminated against the CMSSM for the
  assumed signal and 1 fb$^{-1}$ of LHC data at $\sqrt{s}=14$ TeV. 
  However, mGMSB would not be discriminated on the basis of the kinematic
  end-points alone, and would require further, more detailed investigation.
  The best-fit points of mGMSB and CMSSM look remarkably similar, making
  experimental discrimination at the LHC appear unlikely by any means.
}

\begin{document}
\maketitle
\section{Introduction}

\begin{figure}
  \begin{minipage}{0.99\textwidth}
    \begin{center}
      \epsfig{file=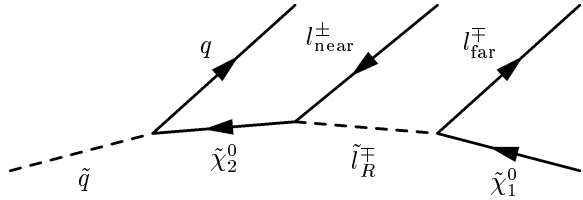}
    \end{center}
  \end{minipage}
  \caption{The golden decay chain $\tilde{q}_L \to \chi_2^0q \left(   \to \tilde{l}^{\pm} l^{\mp} q \right) \to \chi_1^0 l^+ l^- q$}
  \label{fig:goldendecay}
\end{figure}

The Large Hadron Collider (LHC) is currently actively engaged in searches for
new 
physics, including supersymmetry (SUSY). No signal has yet been found, and the
CMS and ATLAS experiments have significantly extended previous exclusion
limits~\cite{daCosta:2011qk,Khachatryan:2011tk}. In the near future, as more
data is collected by the experiments, the observation of a supersymmetric
signal is quite plausible. In the event of a signal, it will be important to
extract as much empirical information as possible about the sparticle
spectrum, since it contains clues about the mechanism of supersymmetry
breaking. We may hope to rule out one mechanism in favour of another. 
One will want to bring all of the data that robustly constrain the
supersymmetry breaking mechanism to bear in order to separate different models
empirically. However, the usual search variables (number of events past
certain cuts or total cross-sections), while perfectly suited to searching for
supersymmetry, are blunt instruments when it comes to measuring supersymmetric
masses in detail: they give only gross information about the overall mass
scale of the supersymmetric particles. Since this is typically described by
some parameter in the SUSY breaking mechanism, such measurements will not tend
to be very good at disentangling models. One needs to measure observables
which are sensitive to the mass spectrum of the sparticles, reasonably
accurate, and robust with respect to experimental systematics such as how well
one has parameterised one's detector. 
Arguably the best examples of such observables come from SUSY cascade decays.
SUSY cascade decay chains give specific kinematics to
the final state particles, and particular kinematic variables have been
shown to contain a wealth of information about the sparticle masses. Maxima
and minima of invariant 
mass distributions, if observed, have several advantages in the inference of
sparticle masses. They can be essentially Standard Model background-free,
particularly if flavour subtracted. Also, although the
shape~\cite{Miller:2005zp} of the 
distributions themselves are subject to significant detector corrections,
which may 
require a lot of integrated luminosity to model well, the end-points of the
distributions are expected to be much less sensitive to such effects. 
For example, the golden decay chain
\begin{equation}
\tilde{q}_L \to q\chi_2^0 \left(   \to \tilde{l}^{\pm} l^{\mp} q \right) \to
\chi_1^0 l^+ l^- q, 
\end{equation}
shown in Figure~\ref{fig:goldendecay} has been shown to be most useful~\cite{Hinchliffe:1996iu}.
The presence of this cascade leads to events with two opposite-sign
same-flavour (OSSF) leptons, jets and missing energy. 
The end-points yield useful information coming from the invariant mass
distributions of the 
di-leptons $m_{ll}$, from the jet and lepton pair $m_{llq}$ and from each
lepton and the jet $m_{lq}$. 
Despite the fact that one obtains highly correlated
mass measurements from such end-points, considering the measurements in
parallel helps discriminating different models of supersymmetry
breaking~\cite{Allanach:2000kt}. 
Even if additional decay chains are identified
in the data, they are not expected to add significant discriminatory power
over the dominant golden chain.
It is by no means guaranteed that the golden
decay chain is present however, for instance it only exists in about a quarter
of the parameter space~\cite{Gjelsten:2004ki} of the constrained minimal
supersymmetric standard model (CMSSM). In the case that the golden chain is
not present, one would use kinematic edge data from all chains that one can
identify. The resulting information is then likely to be less constraining on
the sparticle spectrum than the golden chain. We then view studies assuming
the observation of 
the golden decay chain to be the most optimistic cases as far
as model discrimination goes. If two models cannot be experimentally
discriminated with this assumption, it is extremely unlikely that they will be
discriminated between without the golden cascade. 
The kinematic data have been further combined with
cross-section information 
in order to improve the precision of mass measurements within particular
models with more parameters than the CMSSM~\cite{Lester:2005je}.

Combining the power of the LHC and a linear collider leads to much more
information about the model than is possible from LHC measured kinematic
endpoints, and 
constitutes a significant improvement  on the information
obtained from the LHC alone~\cite{Weiglein:2004hn}. 
Using SUSY signal measurements from both a
linear collider and the LHC in order to measure a large part of the MSSM
spectrum may be possible, allowing checks of unification relations in various
models~\cite{Blair:2000gy,Blair:2002pg,Allanach:2004ud}. The additional
information coming from linear collider data would be ideal to include in
order to discriminate models, but in this paper we restrict ourselves to
potential LHC data, since the linear collider is not yet built. 

Kinematic edge predictions resulting from golden chain decays have been
examined in the literature to see if there could be model discrimination
coming from their measurement. In Ref.~\cite{Allanach:2001qe}, it was seen
whether the ratios of the measurements 
would discriminate the CMSSM, an intermediate-scale string model and a mirage
unification model. The parameters of the models were all scanned over, but no
experimental errors were taken into account. In any case, it was concluded 
that there was no clear separation between the models from using the edge
variables, even for infinitely precise measurements. We go beyond this work by 
examining different models, and by fixing a benchmark model such that we can use
the experimental resolutions estimated by ATLAS, assuming a certain integrated
luminosity. 
In Ref.~\cite{Allanach:2003jw}, the golden decay chain was used in fits to the
CMSSM\@. Hypothetical invariant mass end-points were fit using different
sparticle spectrum calculators in order to examine the differences between them,
quantifying the theoretical error. The best-fit values of each spectrum
calculator were within $95\%$ confidence level (C.L.) limits of each
other, assuming a huge LHC luminosity (300 fb$^{-1}$). A number of other
fitting groups have investigated the effects of LHC data on global fits to the
CMSSM, including the Fittino Collaboration~\cite{Bechtle:2009ty},
SFitter~\cite{Lafaye:2007vs}
and Refs.~\cite{Roszkowski:2009ye,Dreiner:2010gv}. Those works focused more
on the 
constraining power of the LHC data on CMSSM fits. 
In Ref.~\cite{AbdusSalam:2009tr}, current indirect data on $B$ decays,
electroweak observables and the dark matter relic density were combined with
direct sparticle search limits in fits to the CMSSM, mAMSB, LVS and mGMSB
models (to be introduced below) in order to examine whether current data show
any preference for the 
model of supersymmetry breaking. It was found that current indirect data is
too weak to select any of the models. 
On the other hand, end-point data taken from the golden cascade would be
enough to robustly constrain the CMSSM in 1 fb$^{-1}$ of integrated
luminosity, at a particular benchmark point studied by ATLAS, called
SU3~\cite{Roszkowski:2009ye}. 
Such robustness is signalled by prior independence in Bayesian fits,
indicating that the data is sufficiently powerful to constrain the model
hypothesised. 
Ref.~\cite{Fowlie:2011vf} also examined fits from the 
SU3 point golden cascade fits on the CMSSM (with and without including
cosmological data) as well as models with more free
parameters than the CMSSM\@. Model comparison between the
non-universal higgs model, the CMSSM and the CMSSM but with non-universal
gaugino masses was examined using Bayesian techniques. 
Some non-robustness in
the non-CMSSM models with respect to changing the priors was discovered: there
was not 
enough power in the data to properly constrain the models with larger
parameter spaces. 

Since the CMSSM may be robustly constrained by the end-point data, but models
with more parameters may not, in the present paper we answer the following
question: 
Is kinematic edge data from 1 fb$^{-1}$ of the
14 TeV LHC constraining enough to 
allow us to distinguish between simpler different models of supersymmetry
breaking (i.e.\ with fewer parameters in the CMSSM)?
This question will require a numerical statistical analysis: even if it is
clear analytically that a model can be chosen such that its mass spectrum is
close to the CMSSM, the question is: can it be made close {\em enough} in
terms of the errors on the observables to provide a viable fit? 
Conversely, even if two models cannot exactly reproduce the same mass
spectrum, are the errors on the observables small enough such that the two
models are discriminated? 
We shall test robustness by looking for a {\em lack}\/ of prior dependence in
the 
hypothesis testing, and agreement between Bayesian and frequentist inferences.

\subsection{SUSY Breaking Models~\label{sec:sus}}
 
In this subsection, we summarise the alternative hypotheses of SUSY breaking 
that we shall use.
The parameters of the CMSSM are: a flavour
blind SUSY breaking scalar mass
$m_0$, a common gaugino mass $M_{1/2}$, a flavour blind SUSY breaking scalar
trilinear coupling  
$A_0$ and $\tan \beta$, the ratio of the minimal supersymmetric standard model
(MSSM) Higgs vacuum expectation
values (VEVs). Below a grand unification theory (GUT) scale of 
$ M_{GUT} \sim 2 \times 10^{16}$ GeV, the SUSY breaking terms of
different flavours evolve separately to the weak scale.
In anomaly mediated SUSY breaking \cite{Randall:1998uk}
SUSY-breaking is communicated to the
visible sector via the super-Weyl anomaly. 
In its original manifestation, pure anomaly mediation suffers from negative
slepton mass squared parameters, signalling a scalar potential minimum 
inconsistent with a massless photon. 
Minimal AMSB (mAMSB) assumes the existence of
an additional contribution to 
scalar masses $m_0$ at $M_{GUT}$
giving it a total of three 
parameters: the VEV of the auxiliary field in the supergravity multiplet
representing the overall sparticle mass scale, $m_{aux}$, $m_{0}$ and $\tan
\beta$.  
As advertised above, minimal gauge mediated SUSY breaking (mGMSB)~\cite{hep-ph/9801271}
also has three continuous parameters: the overall messenger mass scale,
$M_{mess}$, 
a visible sector soft SUSY-breaking mass scale, $\Lambda$ and $\tan\beta$.
It also contains an additional discrete parameter, namely $N_{mess}$, the
number of SU(5) $5\oplus \bar 5$ representations of mediating fields.  
The example of a moduli mediated model which we consider 
is the Large Volume Scenario (LVS) derived in the context of $IIB$ flux
compactification~\cite{hep-th/0502058, hep-th/0505076,hep-th/0610129,Allanach:2008tu}, whose two extra-SM parameters
can be parametrised by a universal scalar mass $m_0$ and $\tan \beta$.  
At an intermediate scale of $10^{11}$ GeV, the LVS has a universal gaugino
mass $M_{1/2}=\sqrt{3}m_0$ and a universal trilinear scalar coupling
$A_0=-\sqrt{3}m_0$. 

In Section~\ref{sec:kin} following, we detail the predictions of the
golden cascade edges, as well as the expected precision that would come from
LHC measurements. We also specify the SU3 CMSSM benchmark. In
Section~\ref{sec:inf}, we summarise the statistics we shall use to perform
hypothesis testing on the different SUSY breaking models, defining parameter
ranges for the fits.
The results of the hypothesis tests are given in Section~\ref{sec:fit}. We
show that 
mGMSB cannot be discriminated from SU3 by the edge data alone. 
It is then examined in more detail. We sum up and conclude in
Section~\ref{sec:con}. 

\section{Kinematic Edges at SU3 \label{sec:kin}}
The ATLAS collaboration has published a series of studies on reconstructing
SUSY benchmark points in the Supersymmetry section of~\cite{Aad:2009wy}. We
are specifically interested in the study of the CMSSM SU3 benchmark point and
associated mass reconstruction using kinematic end points from golden
cascades. The input parameters
for the SU3 point are shown in
Table~\ref{tab:SU3}. SU3 is a point in the bulk region of the parameter space
with $m_{\chi_1^0} = 118$~GeV and $m_{\tilde{g}} = 720$~GeV. Its
spectrum contains the mass ordering $m_{\chi_1^0} < m_{\tilde{l}} <
m_{\chi_2^0} < m_{\tilde{q}}$ or $m_{\chi_1^0} < m_{\chi_2^0} <
m_{\tilde{l}} < m_{\tilde{q}}$ so that the golden decay chain
is active (in the latter case, the $\chi_2^0$ decay is three-body as the
$\tilde{l}_R$ is off-shell). 
We note that the SU3 point has recently been ruled out by the ATLAS
experiment's jets 
plus zero lepton missing transverse momentum
search~\cite{daCosta:2011qk,Dolan:2011ie}.  
This does not matter for the purposes of the present paper: one must simply bear
in mind that a heavier point will have decreased statistics, and consequently
will require more luminosity to discriminate against other models. 

In the golden decay chain in Fig.~\ref{fig:goldendecay}, one may construct
several Lorentz 
invariant quantities from the four momenta of the visible particles: the quark
and leptons. These are predicted to have various maxima and minima, each
predicted by the theory to be related to the masses of the supersymmetric
particles involved in the cascade decay. We shall now detail this dependence,
which differs depending on whether the $\chi_2^0$ decays through a two body
decay with an on-shell slepton ($m_{\chi_2^0} > m_{\tilde{l}_R}$) or a three
body decay 
($m_{\chi_2^0} < m_{\tilde{l}_R}$). We now detail each case in turn,
collecting the edge predictions from
Refs.~\cite{Allanach:2000kt,Lester:2006cf} for completeness.

\subsection{Prediction of kinematic edges with an on-shell slepton}
One kinematic maximum that we use is 
the di-lepton mass edge. In terms of the sparticle
masses, it is predicted to be
\begin{equation}
{m_{ll}^{\text{edge}}}^2 = \frac{(m_{\chi_2^0}^2-m_{\tilde l}^2) (m_{\tilde
    l}^2 - 
  m_{\chi_1^0}^2)}{m_{\tilde l}^2}. \label{dilept}
\end{equation}
There are two $lq$ edges, in ascending order ${m_{lq(\text{low})}}$ and
${m_{lq(\text{high})}}$, respectively. They are defined to be the maximum
or minimum of various quantities for $m_{\chi_2^0} >
m_{\tilde{l}_R}$:
\begin{eqnarray}
{m_{lq(\text{high})}}&=&\mbox{max} \left[ 
m_{lq}^{{nr}}, m_{lq}^{far}
\right] \label{lqdefs1}\\
{m_{lq(\text{low})}}&=&\mbox{min} \left[ 
{m_{lq}^{{nr}}(\text{max})}, {m_{lq}^{far}(\text{max})}, {m_{lq}'(\text{max})}
\right], \label{lqdefs2}
\end{eqnarray}
where the quantities on the right hand side are defined to be:
\begin{eqnarray}
{m_{lq}^{nr}}^2(\text{max})&=&\frac{(m_{\tilde
    q}^2-m_{\chi_2^0}^2)(m_{\chi_2^0}^2-m_{\tilde 
    l}^2)}{m_{\chi_2^0}^2}, \label{lqnear} \\
{m_{lq}^{far}}^2(\text{max})&=&\frac{(m_{\tilde
    q}^2-m_{\chi_2^0}^2)(m_{\tilde l}^2-m_{\chi_1^0}^2)}{m_{\tilde
    l}^2}, \label{lqfar} \\
{m_{lq}'}^2(\text{max})&=&\frac{(m_{\tilde q}^2 - m_{\chi_2^0}^2)(m_{\tilde l}^2 - m_{\chi_1^0}^2)}
     {2 m_{\tilde l}^2 - m_{\chi_1^0}^2}. \label{lqprime}
\end{eqnarray}
The $llq$ edge is defined as 
\begin{eqnarray}
{m_{llq}^{\text{edge}}}^2 &=& \text{max} \left[
\frac{(m_{\tilde q}^2 - m_{\chi_2^0}^2)(m_{\chi_2^0}^2-m_{\chi_1^0}^2)}{m_{\chi_2^0}^2},\
\frac{(m_{\tilde q}^2 - m_{\tilde l}^2)(m_{\tilde
    l}^2-m_{\chi_1^0}^2)}{m_{\tilde l}^2},\right. \nonumber \\
&&\left. \frac{(m_{\tilde q}^2m_{\tilde l}^2 - m_{\chi_2^0}^2
  m_{\chi_1^0}^2)(m_{\chi_2^0}^2-m_{\tilde l}^2)}{m_{\chi_2^0}^2m_{\tilde
    l}^2} \right]
\end{eqnarray}
unless $m_{\tilde l}^4 < m_{\tilde q}^2 m_{\chi_1^0}^2 < m_{\chi_2^0}^4
$ and $m_{\chi_2^0}^4 m_{\chi_1^0}^2<m_{\tilde q}^2 m_{\tilde l}^4$, in which
case it the right-hand side is equal to $(m_{\tilde q}-m_{\chi_1^0})^2$. 
For the $llq$ {\em threshold}\/ variable, the prediction
is
\begin{eqnarray}
{m_{llq}^{\text{thr}}}^2&=&\frac{1}{4 m_{\tilde l}^2m_{\chi_2^0}^2} \left[
2 m_{\tilde l}^2 (m_{\tilde
  q}^2-m_{\chi_2^0}^2)(m_{\chi_2^0}^2-m_{\chi_1^0}^2)
+\nonumber \right.\\ && \left.
(m_{\tilde
  q}^2+m_{\chi_2^0}^2)(m_{\chi_2^0}^2-m_{\tilde l}^2)
(m_{\tilde l}^2-m_{\chi_1^0}^2)
- \right. \nonumber \\ && \left.
(m_{\tilde
  q}^2-m_{\chi_2^0}^2)
\sqrt{
(m_{\chi_2^0}^2+m_{\tilde l}^2)^2(m_{\tilde l}^2+m_{\chi_1^0}^2)^2
-16 m_{\chi_2^0}^2m_{\tilde l}^4m_{\chi_1^0}^2
}
\right]. \label{thr}
\end{eqnarray}
This edge is the $m_{llq}$ minimum for all events for which
$\frac{1}{\sqrt{2}}\leq m_{ll}/m_{ll}(\text{max})\leq 1$. 
\subsection{Prediction of kinematic edges with three-body $\chi_2^0$ decay}
When $m_{\chi_2^0} < m_{\tilde{l}_R}$, the $\chi_2^0$ decays via a virtual
$\tilde{l}_R$ into leptons and $\chi_1^0$, and in this case the above
Eqs.~\ref{dilept}-\ref{thr} should be altered to the
following: 
\begin{eqnarray}
{m_{ll}^{\text{edge}}}^2&=&(m_{\chi_2^0}-m_{\chi_1^0})^2, \\
{m_{lq(\text{high})}^2}&=& \frac{(m_{\tilde
    q}^2-m_{\chi_2^0}^2)(m_{\chi_2^0}^2-m_{\chi_1^0}^2)}{2 m_{\chi_2^0}^2},\\
{m_{lq(\text{low})}}&=&\frac{m_{lq(\text{high})}}{\sqrt{2}}, \\
{m_{llq}^{\text{edge}}}^2&=& \left\{ \begin{array}{l}
(m_{\tilde q}^2 - m_{\chi_1^0}^2)^2 \text{~if~}
m_{\chi_2^0}^2 > m_{\tilde q} m_{\chi_1^0}, \\
 (m_{\tilde q}^2 - m_{\chi_2^0}^2)(m_{\chi_2^0}^2 - m_{\chi_1^0}^2)/m_{\chi_2^0}^2 \text{~otherwise,}
 \end{array} \right.\\
{m_{llq}^{\text{thr}}}^2&=&\frac{(m_{\chi_2^0}-m_{\chi_1^0})^2}{2} +
\frac{m_{\tilde{q}_L}^2-m_{\chi_2^0}^2}{4 m_{\chi_2^0}^2} \left(
3 m_{\chi_2^0}^2 - m_{\chi_1^0}^2 - 2 m_{\chi_2^0}m_{\chi_1^0} -
\right. \nonumber \\
&& \left. \sqrt{m_{\chi_2^0}^4 + m_{\chi_1^0}^4 + 4 m_{\chi_2^0}m_{\chi_1^0}
(m_{\chi_2^0}^2 + m_{\chi_1^0}^2)-10m_{\chi_2^0}^2m_{\chi_1^0}^2
} \right).
\end{eqnarray}
There is obviously less information than in the case where the slepton is
on-shell, because there are less constraints coming from 4-momentum
conservation. In particular, we see that $m_{{\tilde l}_R}$ does not feature
in the equations, and there is no information on its mass held in the
kinematic edges. 

\subsection{ATLAS reconstruction of the edges}
\begin{table}
\begin{center}
\begin{tabular}[r]{|c|c|c|c|c|c|}\hline
Parameter & $m_0$ & $m_{1/2}$ & $A_0$ & $\tan \beta$ & sgn$\mu$\\ \hline
Value  & 100~GeV & 300~GeV & -300~GeV & $6$ & $+1$\\ \hline
\end{tabular}
\end{center}
\caption{Input parameters of the CMSSM SU3 benchmark
  point. \label{tab:SU3}}
\end{table}

ATLAS have calculated the expected positions of the \mll, \mllqedge, \mllqthr,
\mlqlow~and \mlqhigh~mass distributions. We re-calculate these using the
spectrum obtained for the SU3 point from
\texttt{SOFTSUSY3.1.7}~\cite{Allanach:2001kg}.  
 We take into account the possibility
that $m_{\tilde{l}} > m_{\chi_2^0}$ leading to a three-body decay~\cite{Lester:2006cf}\footnote{The
  presence of the two-body versus the three-body decay can affect the shape of
  the distribution of the di-lepton invariant mass. We do not take this into
  account into our fits, considering only the position of the edge and not its
  shape.}. Since it is not possible to reconstruct the individual squark
masses or flavour, we consider $m_{\tilde{q}_L}$ to be the average of the
masses of the $\tilde{u}_L$ and $\tilde{d}_L$ squarks, as do ATLAS\@. In
Table~\ref{tab:su3edges} we show the positions of the edges as calculated by
ATLAS, and those which we obtain from \texttt{SOFTSUSY3.1.7}. For the di-lepton
edges, the \texttt{SOFTSUSY3.1.7} values are approximately 4~GeV higher than
those given by ATLAS, and for the edges and thresholds involving quarks the
discrepancy is larger, around 20-30~GeV. We have also checked that all the
models 
possess the necessary mass ordering for all edges to exist simultaneously in
at least some part of their parameter space. For instance, in mAMSB this can
be achieved when $m_{aux}/m_0 \sim 10$.  

\begin{table}
\begin{center}
\begin{tabular}[t]{|c|c|c|c|}
\hline
Mass Distribution & ATLAS theory & reconstruction & \texttt{SOFTSUSY3.1.7}\\ \hline
\mll & 100.2 & $99.7 \pm 1.4$ & 103.9  \\ \hline
\mllqedge & 501 & $517 \pm 33.7$ & 532 \\ \hline
\mllqthr & 249 & $265 \pm 23.7$ & 265 \\ \hline
\mlqlow & 325 & $333 \pm 11.7$ & 344 \\ \hline
\mlqhigh & 418 & $445 \pm 19.0$ & 446 \\ \hline 
\end{tabular} \end{center}
\caption{This table shows the position of the endpoints and thresholds for the
  SU3 CMSSM   point in GeV. The column labelled `ATLAS theory' is as predicted
  by   ISAJET7.75~\cite{Paige:2003mg} and used in the experiment's
  simulations.  The simulations of SUSY signal events in 1 fb$^{-1}$ of 14 TeV
  LHC collisions yielded the values marked in the reconstruction column. 
  The final column shows the SU3 values predicted by \texttt{SOFTSUSY3.1.7}.
\label{tab:su3edges}}  
\end{table}
With the SU3 spectrum, ATLAS simulated 1 fb$^{-1}$ of LHC data at
$\sqrt{s}=14$~TeV 
centre of mass energy and simulated the reconstruction of the positions of
the edges and 
thresholds. Full details are available in~\cite{Aad:2009wy}. The results of
this reconstruction are shown in column three of Table~\ref{tab:su3edges},
which shows the central values of the reconstructed edges and an estimate of
the total error which is arrived at by combining in quadrature the estimated
statistical, systematic and jet energy scale (JES) errors. For each edge, we
further 
assume a theoretical error on the {\tt SOFTSUSY3.1.7} prediction of the edge
of half of the difference between  {\tt SOFTSUSY3.1.7} prediction and
the number under the ATLAS theory column of the table. 
We fit the four SUSY breaking models listed 
in Section~\ref{sec:sus} 
to the reconstructed end-points in Table~\ref{tab:su3edges}.  
We have thus neglected the correlations in JES and other systematic errors. 
This should be a reasonable approximation for our purposes, and is
conservative in the sense that including the correlations would actually {\em
  decrease}\/ the total error volume. Thus, if we conclude that two models may
be discriminated by including the errors independently, we may conclude that
the would also be discriminated by including the measurement correlations.

\section{Inference and Fit Details\label{sec:inf}}
Assuming some model hypothesis $H$, 
Bayesian statistics helps update a probability density function
(PDF) $p(\underline m|H)$ of model parameters $\underline m$ with data. The
prior encodes our 
knowledge or 
prejudices about the parameters. Since $p(\underline m|H)$ is a PDF in
$\underline m$,
$\int p(\underline m|H) d\underline m=1$, which defines a normalisation of the prior. One talks of
priors being `flat' in some parameters, but care must be taken to refer to the
measure of such parameters. A prior that is flat between some ranges in a
parameter $m_1$ will not be flat in a parameter $x \equiv \log m_1$, for
example.  
The impact of the data is encoded in the likelihood, or the PDF of obtaining
data set $\underline d$ from model point $\underline m$: $p(\underline
d|\underline m,H) \equiv {\mathcal L}(\underline m)$. 
The likelihood is a function of $\chi^2$, i.e.\ a statistical measure of how
well the data are fit by the model point. One useful quantity is the
posterior: the PDF of
the model parameters $m$ given some observed data $\underline d$ and assuming hypothesis $H$:
$p(\underline m | \underline d, H)$.
Bayes' theorem states that
\begin{equation} p(\underline m|\underline d, H) =
\frac{p(\underline d|\underline m,H)p(\underline m|H)}
{p(\underline d|H)},
\label{eq:bayes}
\end{equation}
where $p(\underline d|H) \equiv \mathcal{Z}$ is the Bayesian evidence, the probability
density of observing data set $d$ integrated over all model parameter space.
The Bayesian evidence is given by: 
\begin{equation}
\mathcal{Z} =
\int{\mathcal{L}(\underline m)p(\underline m|H)}\ d \underline m
\label{eq:3}
\end{equation}
where the integral is over $N$ dimensions of the parameter
space $\underline m$. We note that the evidence depends upon the ranges of
$\underline m$ assumed. 

In order to select between two models $H_{0}$ and $H_{1}$ one needs to compare
their respective posterior 
probabilities given the observed data set $\underline d$, as follows:
\begin{equation}
\frac{p(H_{1}|\underline d)}{p(H_{0}|\underline d)}
=\frac{p(\underline d|H_{1})p(H_{1})}{p(\underline d|
H_{0})p(H_{0})}
=\frac{\mathcal{Z}_1}{\mathcal{Z}_0}\frac{p(H_{1})}{p(H_{0})},
\label{eq:3.1}
\end{equation}
where $p(H_{1})/p(H_{0})$ is the prior probability ratio for the two models, which we set to unity as we adopt the position that no
mechanism of mediation is \textit{a priori} more likely than any other.
 It can be seen from Eq.~\ref{eq:3.1} that
Bayesian model selection revolves around the evaluation of the Bayesian
evidence. As the average of likelihood 
over the prior, the evidence automatically implements Occam's razor. 
A theory
with fewer parameters has a higher prior density since it integrates to
1 over the whole space. Indeed, a theory with the same number of parameters,
but larger \textit{a priori} parameter ranges will have a correspondingly
smaller evidence, for a similar reason, provided both ranges cover the high
likelihood region. 
There is thus a preference for fewer parameters and smaller ranges, unless the
data strongly require there be more. 
Evaluation of the evidence is a computationally intensive task, and specific
algorithms are required to make it practically possible. 
We use the nested sampling approach of~\cite{Skilling} to evaluate the
evidence. A by-product of this approach is that it also produces posterior
inferences. This method is implemented by the {\sc MultiNest} algorithm
of~\cite{Feroz:2007kg,Feroz:2008xx} which we use in this paper.  

The natural logarithm of the ratio of posterior model probabilities
quantifies the level of 
discrimination between two models:
\begin{equation}
\Delta \log \mathcal{Z} = \log \left[ \frac{p(H_{1}|\underline
    d)}{p(H_{0}|\underline d)}\right]
=\log \left[ \frac{\mathcal{Z}_1}{\mathcal{Z}_0}\frac{p(H_{1})}{p(H_{0})}\right].
\label{eq:Jeffreys}
\end{equation}
We summarise the convention we use in this paper in Table~\ref{tab:Jeffreys}.
\begin{table}
\begin{center}
\begin{tabular}{|c|c|c|c|}
\hline
$|\Delta \log \mathcal{Z}|$ & Odds & Probability & Remark \\ 
\hline\hline
$<1.0$ & $\lesssim 3:1$ & $<0.750$ & Inconclusive \\
$1.0$ & $\sim 3:1$ & $0.750$ & Weak Evidence \\
$2.5$ & $\sim 12:1$ & $0.923$ & Moderate Evidence \\
$5.0$ & $\sim 150:1$ & $0.993$ & Strong Evidence \\ \hline
\end{tabular}
\end{center}
\caption{The Jeffreys' scale of hypothesis testing. Here the `$\log$'
  represents the natural logarithm. \label{tab:Jeffreys}} 
\end{table}

In Bayesian model selection the results will always depend to some extent on the priors. Rather than seeking a unique `right'
prior, one should check the 
independence of conclusions with respect to a reasonable variation of the
priors. Such a 
sensitivity analysis is required to 
ensure that the resulting model comparison is not overly dependent on a particular choice of prior and the
associated metric in parameter space, which controls the value of the integral involved in the computation of the
Bayesian evidence.
Prior dependence has been studied in the CMSSM fitted to indirect data in~\cite{Trotta:2008bp}, where it was demonstrated that the indirect data was not constraining enough to allow a prior-independent determination of the preferred regions of the parameter space. Prior dependence in parameter estimation was also treated in~\cite{Allanach:2006jc,Allanach:2007qk}, and in evidence evaluation in~\cite{AbdusSalam:2009tr,AbdusSalam:2009qd}.

We have considered two different prior PDFs in
this analysis. The first is the standard ``linear prior'' where 
$p(m_1)=p(m_2) $
for $m_{1,2}$  being
two different points in the parameter space of one of
the models under consideration. 
We shall contrast the results with linear priors versus those with log priors:
each parameter $m$ with dimensions of mass has a prior whose distribution is
flat in $\log (m)$, except for $A_0$ in the CMSSM\@. 
$A_0=0$ requires a different
treatment because of the singularity at 0 in $\log A_0$: we choose a prior
that is flat in  $\log (|A_0| + C)$. For this particular study, we
pick $C=60$ GeV, but the results are not at all sensitive to the value chosen
(indeed, we shall see that they are not sensitive to the choice of log or flat
priors - a much larger change).

Before proceeding, we specify the parameter ranges over which we
sample for the different models. 
We consider only the positive sign of $\mu$, as it is well known that the
kinematical edges we consider do not have the power to distinguish the sign of
$\mu$. It is unlikely that the LHC will have enough data to distinguish
different signs of $\mu$: given current search constraints where soft SUSY
breaking terms are expected to be heavy, the sign of $\mu$ may only have a
fairly small effect on aspects of the spectrum. It affects heavier chargino and
neutralino masses and mixings, and the third family sfermion mixings, all of
which will be difficult to measure accurately at the LHC (but which may well
be accurately measured at a future linear collider). 
The ranges over which we vary the continuous
model parameters are shown in Table~\ref{tab:ranges1}. 
\begin{table}
\begin{center}
\begin{tabular}{|c|c|} \hline
CMSSM & mAMSB \\ \hline
 $1\mbox{~GeV}\leq m_0 \leq 2$ TeV & $1\mbox{~GeV}\leq m_0 \leq 2$ TeV \\
$60\mbox{~GeV} \leq m_{1/2} \leq 2$ TeV & $ 20\mbox{~TeV} \leq m_{aux} \leq 100\mbox{~TeV}$  \\
$-4\mbox{~TeV} \leq A_0 \leq 4\mbox{~TeV}$ &   \\ \hline
\hline
mGMSB & LVS \\ \hline 
 $10^4\mbox{~GeV} \leq \Lambda \leq 10^6\mbox{~GeV}$  & $1\mbox{~GeV}\leq m_0\leq 2\mbox{~TeV}$ \\ 
$10^5\mbox{~GeV} \leq M_{mess} \leq 10^{14}\mbox{~GeV}$ & \\ \hline
\end{tabular}
\end{center}
\caption{Ranges for the parameters in mGMSB and the Large Volume Scenario. In
  mGMSB we also vary the discrete parameter $N_{mess}$ between 1 and 8.
For all models, $2\leq \tan \beta \leq 62$.\label{tab:ranges1}}
\end{table}

We bound $\tan \beta$ from below by 2, as values lower than this are in
contravention of LEP2 Higgs searches, and from above by 62, since such large
values lead to non-perturbative Yukawa couplings below the GUT scale and
calculability is lost. 
In mGMSB the discrete parameter $N_{mess}$, the number of
messenger multiplets, is varied between 1 and 8. Higher values of
$N_{mess}$ lead to problems with perturbativity of gauge interactions at the
GUT scale \cite{hep-ph/9801271}. We wish to avoid possible contributions from gravity mediation in our GMSB fits. Gravity mediated contributions will always be present and of order $F/M_{Pl}$, where $\sqrt{F}$ is the supersymmetry breaking scale, and we require these contributions to the soft masses to be less than 1~GeV. This implies a maximum value of $F$ of around $10^{19}$~GeV. Since the mass scale $\Lambda = F/M_{mess} \sim 10^5$~GeV, we restrict $M_{mess}$ to be less than $10^{14}$~GeV.
In the CMSSM the unification scale is the standard GUT scale $M_{GUT} \approx 2\times 10^{16}$GeV, while for 
the LVS the soft terms are defined at the intermediate string  scale $m_s \approx 10^{11}$GeV as in~\cite{Allanach:2008tu}.

The
constraints we use are all shown in Table~\ref{tab:su3edges}. 
We treat the measurements $D_i$ of the observables as independent.
 We also assume Gaussian errors on all
measurements. 
The pull of observable $i$ is 
calculated by
\begin{equation}
s_i = \frac{|c_i-p_i|}{\sigma_i},
\end{equation}
where 
$c_i$ is the central
experimental value of observable $i$, $p_i$ is the prediction of it by the
model point and hypothesis assumed and
$\sigma_i$ is   the standard deviation incorporating both experimental and
theoretical uncertainties, added in quadrature. 
The pull is a measure of how far the prediction is from the central
experimental value in comparison to the error. In the limit of large
statistics, where the experimental measurements have Gaussian probability
distributions, $\chi^2=\sum_i s_i^2$ follows a well-known (`$\chi^2$')
distribution. 
The log likelihood 
of a prediction $p_i$ of an observable $i$ is given by
\begin{equation}
\log \mathcal{L}_i = -\frac{s^2_i}{2} -\frac{1}{2}\log(2\pi) - \log(\sigma_i)
\end{equation}
The combined log likelihood is the sum of the individual log likelihoods,
\begin{equation}
\log\mathcal{L}^{tot} = \sum_i \log\mathcal{L}_i. \label{lik}
\end{equation}
We do not use any indirect observables in this article. If an edge or
threshold is not present 
due to the mass ordering in the spectrum, the likelihood of that point is set
to zero. Eq.~\ref{lik} amounts to assuming that the measurements of each
end-point are independent. This is not strictly true: jet-energy scale errors,
for instance, will tend to correlate \mllqedge, \mllqthr, \mlqlow~and
\mlqhigh, for instance. However, this is not expected to be a large effect, and
neglecting the resulting correlation should yield a reasonable approximation. 
Correlations between the sparticle masses coming from the measurements are
automatically taken into account by Eqs.~\ref{dilept}-\ref{thr}.

Aside from the Bayesian evidence, we shall evaluate the comparative quality of
fit of each model via the $p-$value of their best-fit points. For a given
model, the best-fit point in parameter space is defined to be the one
with the lowest $\chi^2$. 
The $p-$value is constructed as follows: 
it is the probability of obtaining
$\chi^2$ at least 
as large as the one actually observed $\chi^2_o$, assuming the best-fit point
of the hypothesised model: 
\begin{equation}
p=\int_{\chi^2_o}^\infty \frac{1}{2^{k/2} \Gamma(k/2)} x^{k/2-1} e^{-x/2}
\ dx,
\label{pvalue}
\end{equation}
where $k$ is the number of degrees of freedom: the number of observables minus
the number of parameters in the model. $p-$values do not depend upon 
priors. 
However, in common problems, the interpretation
of the $p-$value is problematic because of the identification of the number of
degrees of freedom. One could always add additional observables that are
insensitive to the value of the model parameters at the best-fit point,
changing the value of $p$, for instance. Also, the presence of physical
boundaries may spoil the interpretation of $p$ as calculated in
Eq.~\ref{pvalue}~\cite{Bridges:2010de}. Nevertheless, we use $p-$values as a
qualitative estimator of the overall quality of the fit in each case: a small
$p-$value indicates the fact that the model is not able to fit the data well,
and a $p-$value closer to unity indicates that the model may fit it.
We calculate the $p-$value
by minimising the $\chi^2$ function using the minimiser
{\tt MINUIT}~\cite{James:1975dr} (a
particular configuration of {\sc MultiNest} has also shown to be able to
perform this task~\cite{Feroz:2011bj}).
We use
the point sampled by {\sc MultiNest} during the evidence calculation with the
highest likelihood as a starting 
seed. 

Since we are assuming that the LHC measurements discover missing transverse
momentum like signals, which yield SUSY signals leading to the endpoints
detailed in Table~\ref{tab:su3edges}, we require a neutral MSSM particle that
is stable, at least on time-scales required for it to traverse an LHC
detector. 
Therefore, in mAMSB, the CMSSM and the LVS the neutralino must be the lightest
supersymmetric particle (LSP), or
else we set the likelihood to zero. 
In mGMSB the gravitino $\tilde{G}$ is the LSP, and the collider signatures are
to a 
large part determined by the identity of the next-to-lightest supersymmetric
sparticle (NLSP). If the stau is the NLSP we reject the point, assigning it a
zero likelihood. If the neutralino is the NLSP we consider its decay
length. If the (bino-like) neutralino decays inside the detector, then the
classic di-photon and missing transverse energy of low scale mGMSB is realised,
in contradiction to the signals that we assume from Ref.~\cite{Aad:2009wy}. We
therefore ensure that in mGMSB the NLSP is the neutralino and that it is
stable on detector time-scales. Specifically, we calculate the decay length of
the neutralino according to~\cite{Ambrosanio:1997rv}, where 
\begin{equation}
L_{\mbox{decay}} =  \frac{1}{\kappa_{\gamma}} \left( \frac{100~\mbox{GeV}}{m_{NLSP}} \right)^5 
\left(  \frac{\Lambda}{100~\mbox{TeV}}  \right)^2 
\left(  \frac{M_{mess}}{100~\mbox{TeV}}  \right)^2   10^{-4}~\mbox{m}. 
\end{equation}
where $\kappa_{\gamma}$ is the photino component of the neutralino, since in mGMSB the neutralino NLSP
is predominantly photino-like. 
If the decay length is less than ten
metres we reject the point. 
We also apply some simple direct search bounds, adapted from~\cite{Nakamura:2010zzi,AbdusSalam:2009tr,Abel:2009ve}. If a sparticle
mass falls these bounds, the corresponding point is assigned a zero likelihood.

To calculate the MSSM spectrum we use \texttt{SOFTSUSY3.1.7}
which calculates the spectrum of the CMSSM, mAMSB and mGMSB\@. By modifying the
unification scale from $M_{GUT}$ to $m_{string}\sim 10^{11}$GeV and by not
enforcing gauge
coupling unification, \texttt{SOFTSUSY3.1.7} can also provide the spectrum
in the LVS case. Parameter space points which do not break electroweak 
symmetry correctly or have tachyonic sparticles are assigned zero likelihood. 
However, this disallowed part of parameter space {\em is}\/ included in our
calculation of the prior volume and so will consequently reduce the evidence.
Points which have a charged LSP are
rejected\footnote{Due to the small neutralino-chargino splitting in mAMSB we
  must
reject any points that would violate the long-lived charged stable particle
bounds from Tevatron, which requires $\Delta m= m_{\chi_1^+}-m_{\chi_1^0} >
50$ MeV. In practice, we find that this bound does not constrain the mAMSB
parameter space
since mAMSB predicts larger splittings~\protect\cite{hep-ph/9904378}.}. 

\section{Fits to Edge Data \label{sec:fit}}
\subsection{Hypothesis Testing}
The Bayesian evidence values  calculated for the different models and priors are
shown in Table~\ref{tab:bayesresults}. Although there is a small dependence of
the evidence upon the prior, there is 
a much larger difference between the evidences of some of the models and so we
may expect to reliably discriminate between them on that basis. 
One would strongly discriminate against LVS
and mAMSB in favour of SU3 based either
on the Jeffreys' scale of Bayesian evidence differences or on the $p-$values. 
However,
we see that we would not discriminate between mGMSB and the
CMSSM using the evidence\@. 
\begin{table}
\begin{center}
\begin{tabular}[t]{|c||c|c|c|}
\hline
Model & log${\cal Z}$(linear) & log${\cal Z}$(logarithmic)  & $p-$value\\ \hline
CMSSM & -28.1 & -25.1 & 0.64   \\ \hline
mGMSB & -27.1 & -25.8 & 0.83 \\ \hline
mAMSB & -55.7 & -54.1 & $<10^{-10}$\\ \hline
LVS & -47.0 & -47.0 & $1.4 \times 10^{-9}$ \\ \hline
\end{tabular}
\end{center}
\caption{Hypothesis testing statistics for the different models. The columns
  labelled $Z$ show the Bayesian evidence for either linear or logarithmic
  priors.  
The error on each entry of the Bayesian log${\cal Z}$ delivered by {\sc
  MultiNest} 
is $\pm 0.1$. \label{tab:bayesresults}}   
\end{table}
Reassuringly, the $p-$values point in the same direction as the Bayesian
evidence: mAMSB and LVS would be discriminated against, but mGMSB and the
CMSSM could not be distinguished on the basis of edge data alone. 
The agreement of the interpretation of the naive frequentist ($p-$value) and
Bayesian (evidence) measures of hypothesis test is another signal that the
fits are fairly robust, together with their approximate prior independence. 

\subsection{Best Fit Points}

The best-fit points along with their $\chi^2$ values divided by the
number of 
degrees of freedom ($\chi^2/\mbox{d.o.f}$) and the associated $p-$value are
shown in Table~\ref{tab:bestfit}.
\begin{table}
\begin{center}
\begin{tabular}[t]{|c||c|c|c|}\hline
Model & parameters & $\chi^2/\mbox{d.o.f}$& $p-$value \\ \hline
CMSSM & $m_0=92.1$ GeV, $m_{1/2}=300.6$ GeV  & $0.22/1$ & 0.64\\
 & $A_0=984$ GeV, $\tan \beta=12.3$& & \\ \hline
mAMSB & $m_{aux}=28.46$ TeV, $m_0=255.5$ GeV & $52/2$ & $<10^{-10}$\\
      & $\tan \beta=22.4$ & & \\ \hline
mGMSB & $M_{mess}=1.0\ 10^{14}$ GeV, $\Lambda=1.78\ 10^4$ GeV & 0.36/2 & 0.83
\\
& $N_5=5$, $\tan\beta=22.2$   & & \\ \hline
LVS & $m_0=359$ GeV, $\tan \beta=4.75$ & 44.2/3 & $1.4 \times 10^{-9}$ \\ \hline
\end{tabular}
\end{center}
\caption{Best-fit points (defined as having the highest likelihood) for each
  model, along with the associated value of $\chi^2/\mbox{d.o.f}$ and
  $p$-value. We have assumed that $\mu>0$ for each point. 
\label{tab:bestfit}}
\end{table}
The table illustrates that {\tt SOFTSUSY3.1.7} is able to fit the $\mu>0$
CMSSM to the 
assumed edge variables extremely well, despite the fact that they were
produced by a different SUSY spectrum calculator. This is implied by the
statement that there are only small differences in the masses of sparticles
appearing in the golden decay chain between the spectrum calculators
anyway, as Ref.~\cite{Allanach:2003jw} shows. Performing another fit for
$\mu<0$, we confirm our earlier assertion that the edges we study are not
sensitive to the sign of $\mu$, obtaining a total $\chi^2$ of 0.14 and a
$p-$value of 0.71. Similar fits are obtained for the other models under study
for $\mu<0$ as for $\mu>0$, and so we simply show results of the fits 
for $\mu>0$. Non-LHC data may separate the two signs of $\mu$: famously, the
anomalous magnetic moment of the muon is sensitive to it (and prefers $\mu>0$
in the CMSSM). Also, linear collider measurements of neutralinos and charginos
may accurately constrain all of the parameters appearing in their mixing
matrices, including $\mu$~\cite{Desch:2003vw}.
While we display only the absolute best-fit point in the table for
mGMSB, 
there are in fact best-fit points for $N_5=3,4$ and $6$ which have $p-$values
larger than 0.05, indicating that one would not necessarily discriminate 
against mGMSB with these values of $N_5$ either. 

\begin{figure}
  \begin{minipage}{0.99\textwidth}
    \begin{center}
      \includegraphics[width=0.8\textwidth]{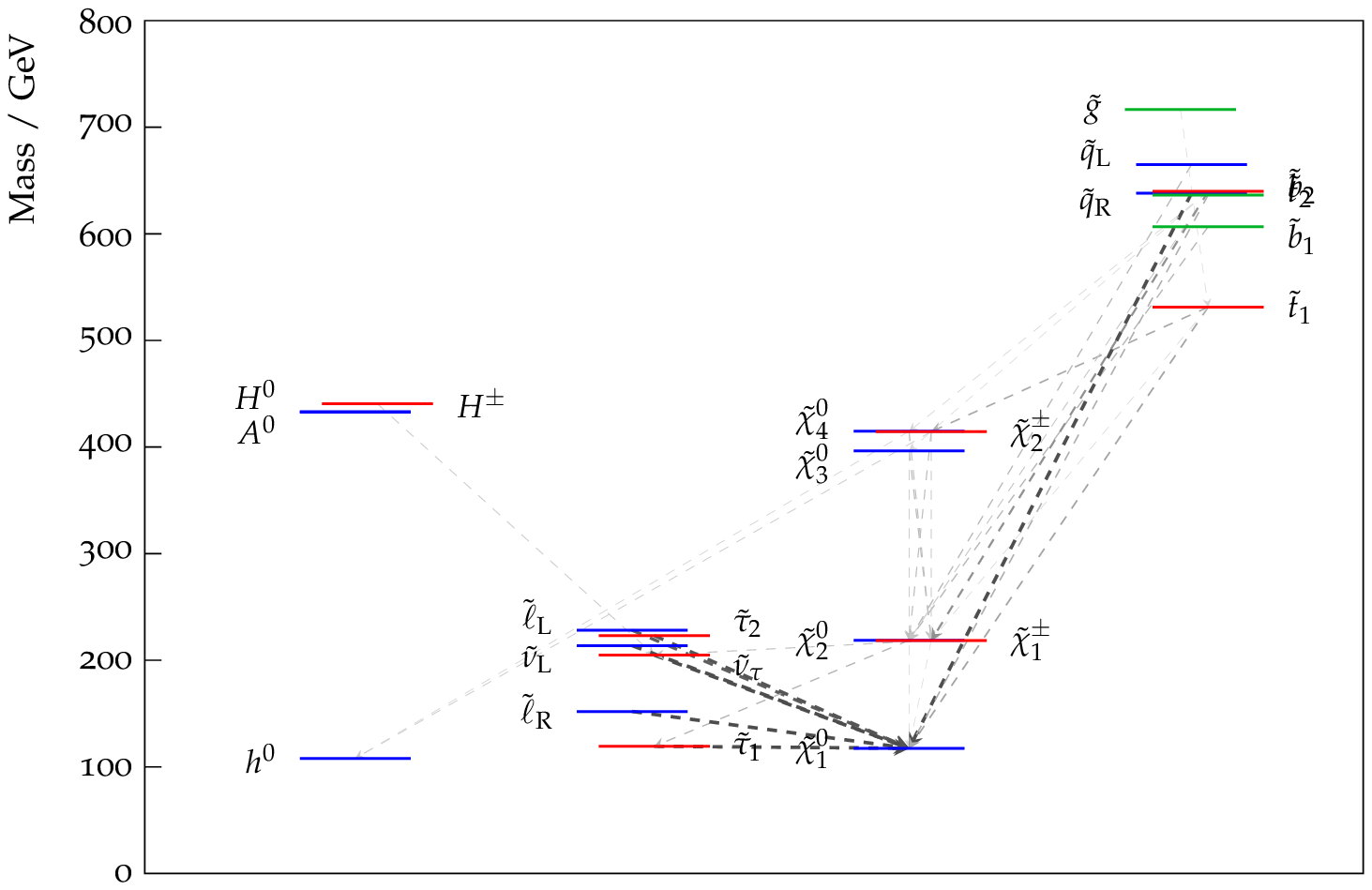}
      \includegraphics[width=0.8\textwidth]{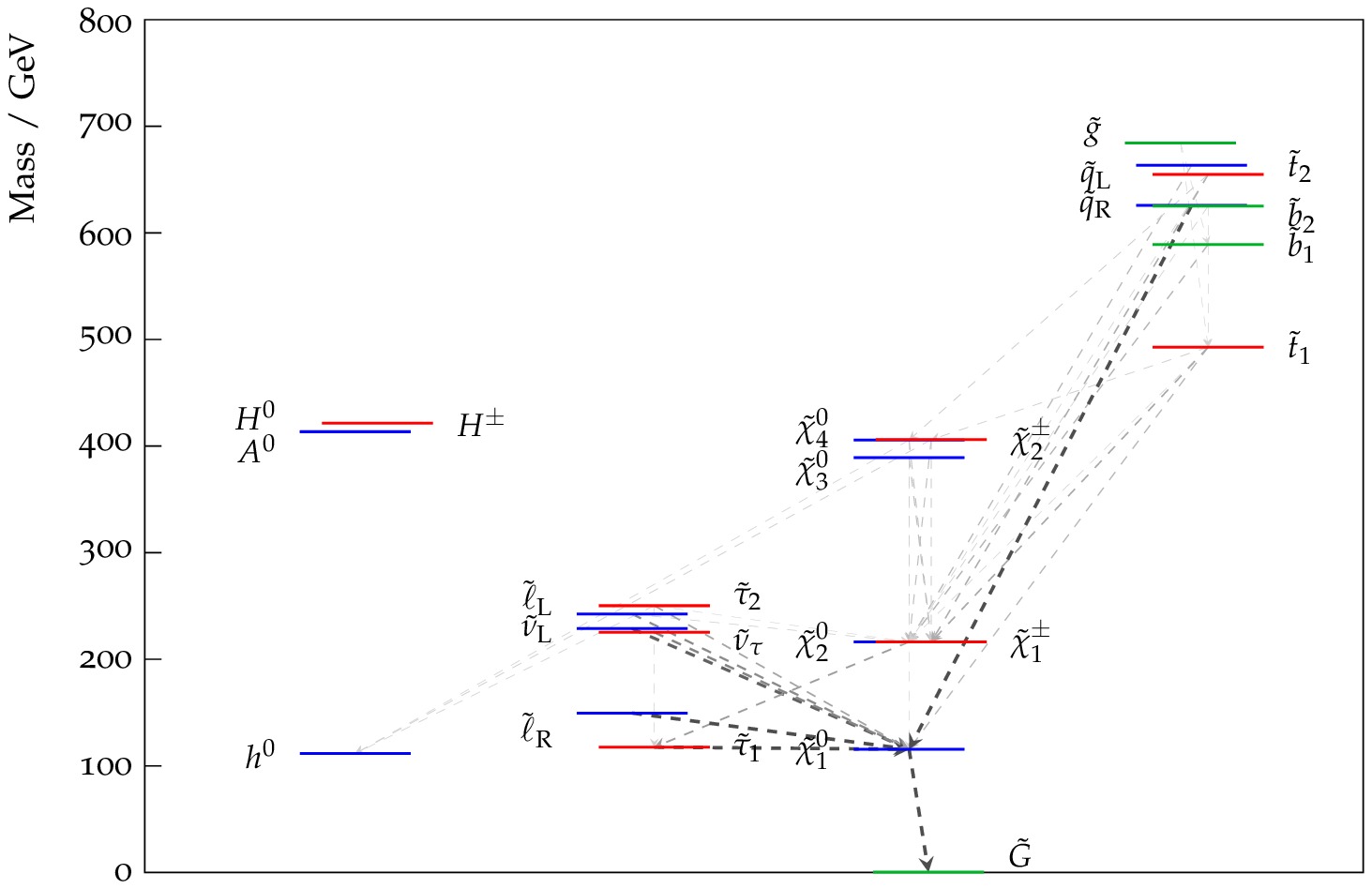}
    \end{center}
  \end{minipage}
\caption{Spectra and decays in the best-fit points of the CMSSM (top panel)
  and mGMSB (bottom panel). Here, the super partners are displayed with tildes,
  unlike in the rest of the paper. 
\label{fig:spectra}}
\end{figure}
We plot the spectra of the CMSSM and mGMSB best-fit points in
Fig.~\ref{fig:spectra}. The decays were calculated with {\tt SDECAY 1.3b}~\cite{Muhlleitner:2003vg}, and
we display only those decays whose branching ratios  are higher than 10$\%$. 
The figure shows that the two best-fit spectra and decays are remarkably
similar, and could prove difficult to discriminate. 
Although the heavier third generation squarks are somewhat heavier in mGMSB,
they may be difficult to access experimentally because decays to them from the
gluino are phase-space suppressed. Although, in the mGMSB panel, the decay of
$\chi_1^0$ to 
gravitino (ejecting a photon) is shown, the neutralinos are actually
quasi-stable and so this decay will not show up in the experiment. We find that the decay length of the neutralino for the best-fit point is about 12.5AU, due to the very high messenger scale.
The splitting between gluino and first two generations of squark (denoted
${\tilde q}_L$
and ${\tilde q}_R$ respectively, in the figure) are smaller for mGMSB, which could potentially make one
of the jets from gluino decay softer, so there potentially could be a
potential discriminator in the hardness of this jet, or indeed the
multiplicity from gluino decays, if the jet is too soft to make it past jet
cuts. A feasibility study of experimental separation between these two models
would require a detailed study, and is beyond the scope of this paper.

\begin{figure}
\begin{center}
\includegraphics[width=0.4\textwidth,angle=270]{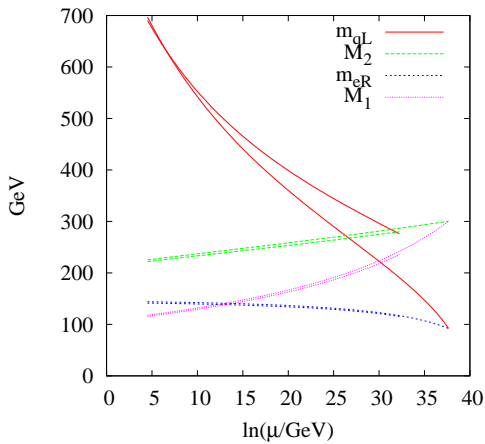}
\end{center}
\caption{Renormalisation of CMSSM and mGMSB best-fit points. We show the most
  relevant $\overline{DR}$ mass parameters as a function of the
  renormalisation scale $\mu$ for each model. The CMSSM model curves continue
  to $\ln (\mu/\mbox{GeV}) \approx 37$, whereas the mGMSB model curves
  terminate at $\ln (\mu/\mbox{GeV}) \approx 32$. \label{fig:run}}
\end{figure}
What leads to the similarities between the mGMSB and CMSSM best-fit points'
spectra? In 
the CMSSM the soft-terms run from the GUT scale, while in mGMSB they run from
the messenger scale $M_{mess}$. We observe that the messenger scale of the
mGMSB best-fit point is as close as possible to the GUT scale given the range
assumed in Table~\ref{tab:ranges1}, $M_{mess}
=1\times 10^{14}$~GeV. Working to one-loop order, since the ratio of each MSSM
group's gaugino 
mass $M_i$ to its gauge coupling squared $g_i^2$ does not run,
if there exists 
a renormalisation scale $\mu=\mu_0$ for which
\begin{equation}
\frac{M_3(\mu)}{g_3(\mu)^2}  =  \frac{M_3(\mu)}{g_2(\mu)^2}  = 
 \frac{M_3(\mu)}{g_1(\mu)^2}, \label{gc}
\end{equation} 
then Eq.~\ref{gc} applies for any $\mu$, in particular at the weak scale. 
In the CMSSM, Eq.~\ref{gc} is satisfied because
$M_3(M_{GUT})=M_2(M_{GUT})=M_1(M_{GUT})$ as well as 
$g_3(M_{GUT})=g_2(M_{GUT})=g_1(M_{GUT})$, whereas the mGMSB soft SUSY breaking
boundary conditions are $M_i(M_{mess})=N_5 \Lambda g_i^2(M_{mess}) f/ (16
\pi^2)$~\cite{Ambrosanio:1997rv}, where 
$f$ is a dimensionless number depending upon parameter space (but not on the
gauge group $i$). The mGMSB gaugino masses thus 
explicitly satisfy Eq.~\ref{gc} in a different way to the CMSSM at
$\mu=M_{mess}$.  
Numerically, substituting $\mu=M_Z$ into Eq.~\ref{gc} leads to the approximate
pattern $M_3:M_2:M_1 \sim 6:2:1$ for the weak-scale gaugino masses, which
applies to 
both mGMSB and the CMSSM\@.

The high-scale scalar mass boundary conditions in mGMSB have more complicated
expressions than in 
the CMSSM, as they depend on the quadratic Casimir operators and $g_i$. They
are not universal at the GUT scale. We find that 
the SUSY breaking right-handed slepton mass parameter for the best-fit mGMSB
point at the GUT scale is
92.8~GeV, close to the CMSSM value of 92.4~GeV. The left-handed
slepton mass parameters are somewhat larger as they are charged under $SU(2)$, but at the
weak scale it is the right-handed sleptons which are lightest and whose mass
parameters 
we use to calculate the edge positions. This difference therefore does not
affect the quality of the fit. The mGMSB squark masses at the messenger scale
are  significantly
different to the CMSSM squark masses which are given by $m_0$. However, during
the renormalisation group running the squark masses are renormalised by the
contributions from 
the gluino, and thus at the low scale the squark masses for both model points
are similar to the gluino mass. Finally, the trilinear $A$-terms differ
for the CMSSM and mGMSB best-fit points, but they affect the end-points 
by less than 1$\%$. 
We display the renormalisation of the most relevant mass parameters in
Fig.~\ref{fig:run}. Since $\chi_1^0$ and $\chi_2^0$ are
approximately bino and 
wino-dominated respectively, tuning $\Lambda$ allows mGMSB to match both
gaugino masses to the ones required by our benchmark CMSSM point in the $2:1$
ratio that applies to both models. The other messenger scale scalar masses are
fixed, but we may then tune $M_{mess}$ to get one of them (say, $m_{{\tilde
    e}_R}(M_{mess})$) to match with its equivalent value in the
CMSSM benchmark. The other mass (in this case $m_{{\tilde q}_R}(M_{mess})$)
is then predicted by mGMSB, and must renormalise (within an accuracy dictated
by the measurement errors) to the tree-level value in the CMSSM benchmark
model. We see from Fig.~\ref{fig:run}, that this is indeed the case.

\begin{figure}
  \begin{minipage}{0.99\textwidth}
    \begin{center}
      \includegraphics[width=0.48\textwidth]{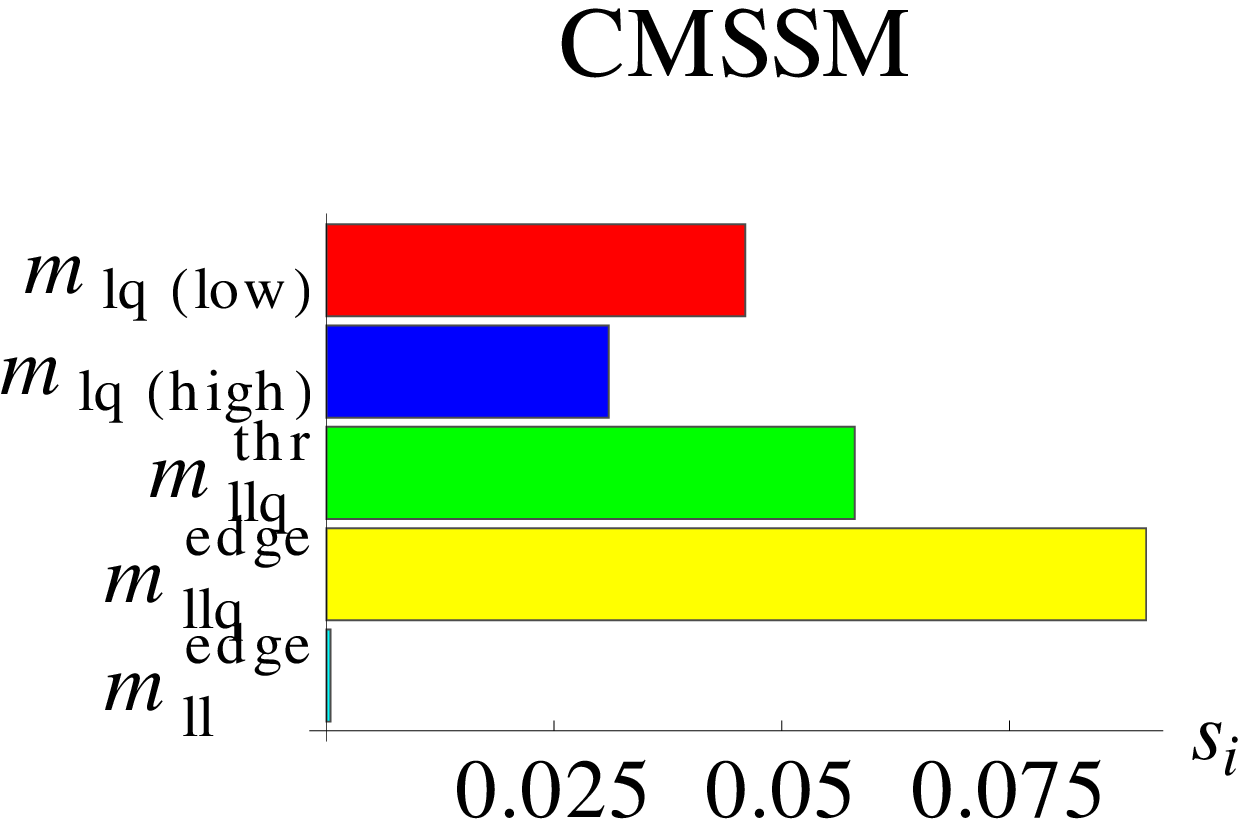}
      \includegraphics[width=0.48\textwidth]{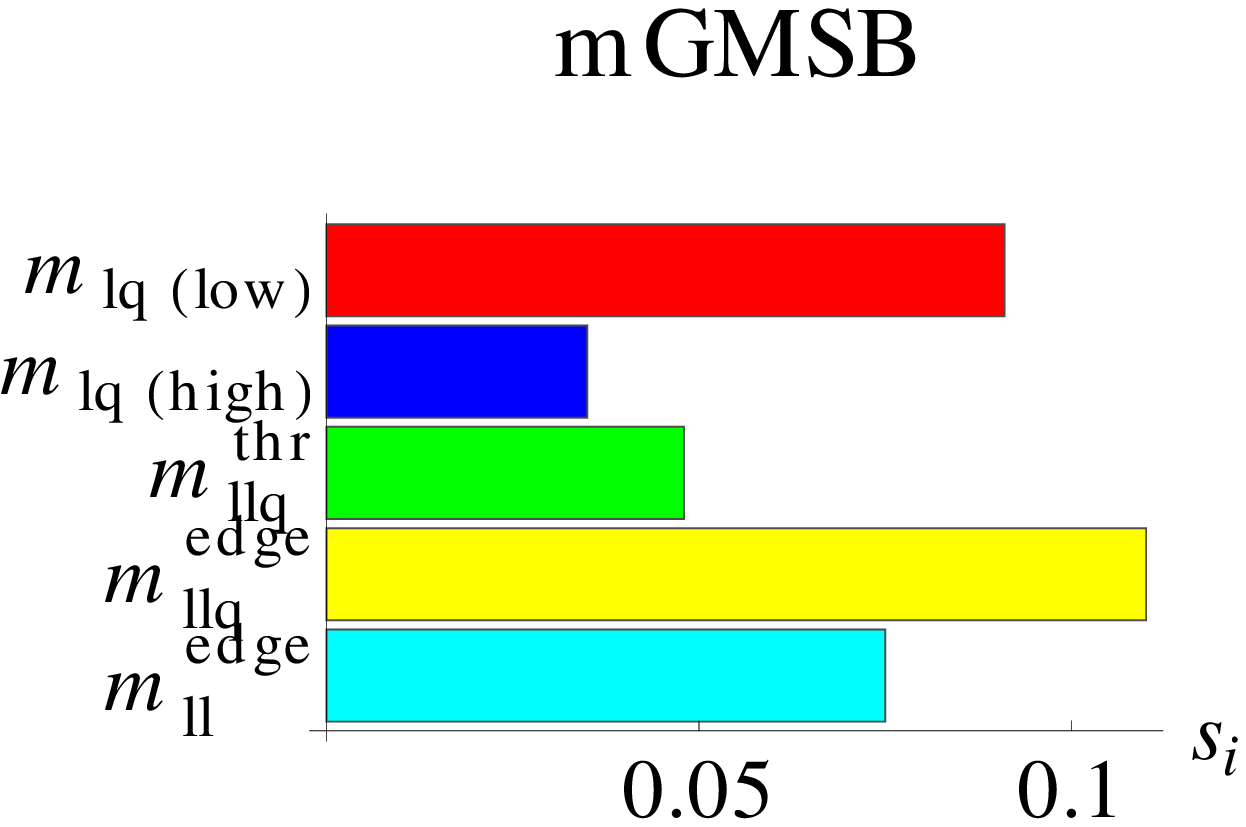}
    \end{center}
  \end{minipage}
\caption{Pulls in the best-fit points of the CMSSM and mGMSB\@.
\label{fig:pulls}}
\end{figure}
The pulls from each observable $s_i$ are displayed in
Fig.~\ref{fig:pulls} for the 
best-fit mGMSB and CMSSM models. 
We see a similar pattern for each of the observables except for \mll, which is
larger for mGMSB\@. However, it is clear that each of the observables is
well-fit by each best-fit model, with no one observable dominating the
$\chi^2$. Note that, even though mGMSB has a higher value of $\chi^2$, it has
a slightly higher $p-$value because it has less free continuous parameters,
and therefore a larger number of degrees of freedom.

We note that the edge information is not the only information one would
collect about the models to use to discriminate them. Before sufficient
statistics have been collected to 
constrain the kinematic edges, we would have rate data on the number of signal
events passing cuts in missing transverse momentum type searches. The production
rates of supersymmetric particles at the LHC typically dominantly depend upon
the squark and gluino masses, since these are the strongly interacting
particles with the largest direct production cross sections. They then decay
in various ways into different channels. The rates 
for the individual channels do have a complicated dependence on the detailed
MSSM parameters, but still: all channels are proportional to the total SUSY
production cross section, which is a function of squark and
gluino masses only, to a good approximation. Therefore the total SUSY
production cross-section is a function 
of squark and gluino masses, and we compare them at the best-fit points
of the CMSSM and mGMSB models in Table~\ref{tab:mass}. We also show the total
next-to-leading order 
SUSY production cross-section as calculated by {\tt
  PROSPINO}~\cite{Beenakker:1996ch}. This is the cross-section {\em without
  cuts or acceptance corrections}, so the measurable cross-section will be
some factor times smaller (around 30 in some examples). 
\begin{table}
\begin{center}
\begin{tabular}[t]{|c||c|c|c|}\hline
               & CMSSM & mGMSB & LVS \\ \hline
$m_{\tilde g}$/GeV & 716   & 686   & 1116\\
$m_{\tilde q}$/GeV & 662   & 662   & 1019\\ 
$\sigma_{NLO}$/pb  & 22 & 25 & 1.7\\ 
\hline
\end{tabular}
\end{center}
\caption{Mass spectra and total next-to-leading order SUSY 14 TeV LHC production
  cross-section $\sigma_{NLO}$ of the best-fit
  points for the CMSSM, mGMSB and LVS models in GeV. $m_{\tilde q}$ is an
  averaged first family squark mass.\label{tab:mass}}
\end{table}
We see from Table~\ref{tab:mass} that the CMSSM and mGMSB have similar squark
and gluino 
masses, resulting in a similar total SUSY LHC production cross-section. Thus,
the models would likely require other more detailed empirical information to
tell them 
apart. We have shown the gluino and squark masses obtained in the LVS best-fit
model, because they are not yet ruled out at 95$\%$ confidence
level by 165 pb$^{-1}$ of LHC data~\cite{atlasconf}, unlike the best fit mGMSB
and CMSSM points. If we scaled up the masses of all sparticles at the mGMSB
and CMSSM points so that squarks and gluino masses are similar to the LVS
best-fit point, we would have a total SUSY cross-section of around $1/10$th of
the value that SU3 has. If the number of events past cuts just scaled with the
total SUSY cross-section, we would then expect to require 10 fb$^{-1}$ of LHC
data in order to achieve similar fractional precisions on the end-points as
the ones assumed in the present paper. Of course, a dedicated 
simulation of LHC collisions would be required to calculate this number more
exactly and to verify that  for heavier spectra mGMSB is indeed able to emulate
the CMSSM spectrum.

\subsection{Posterior Distributions for CMSSM and mGMSB}

\begin{figure}
\begin{center}
\onegraphwcntrs{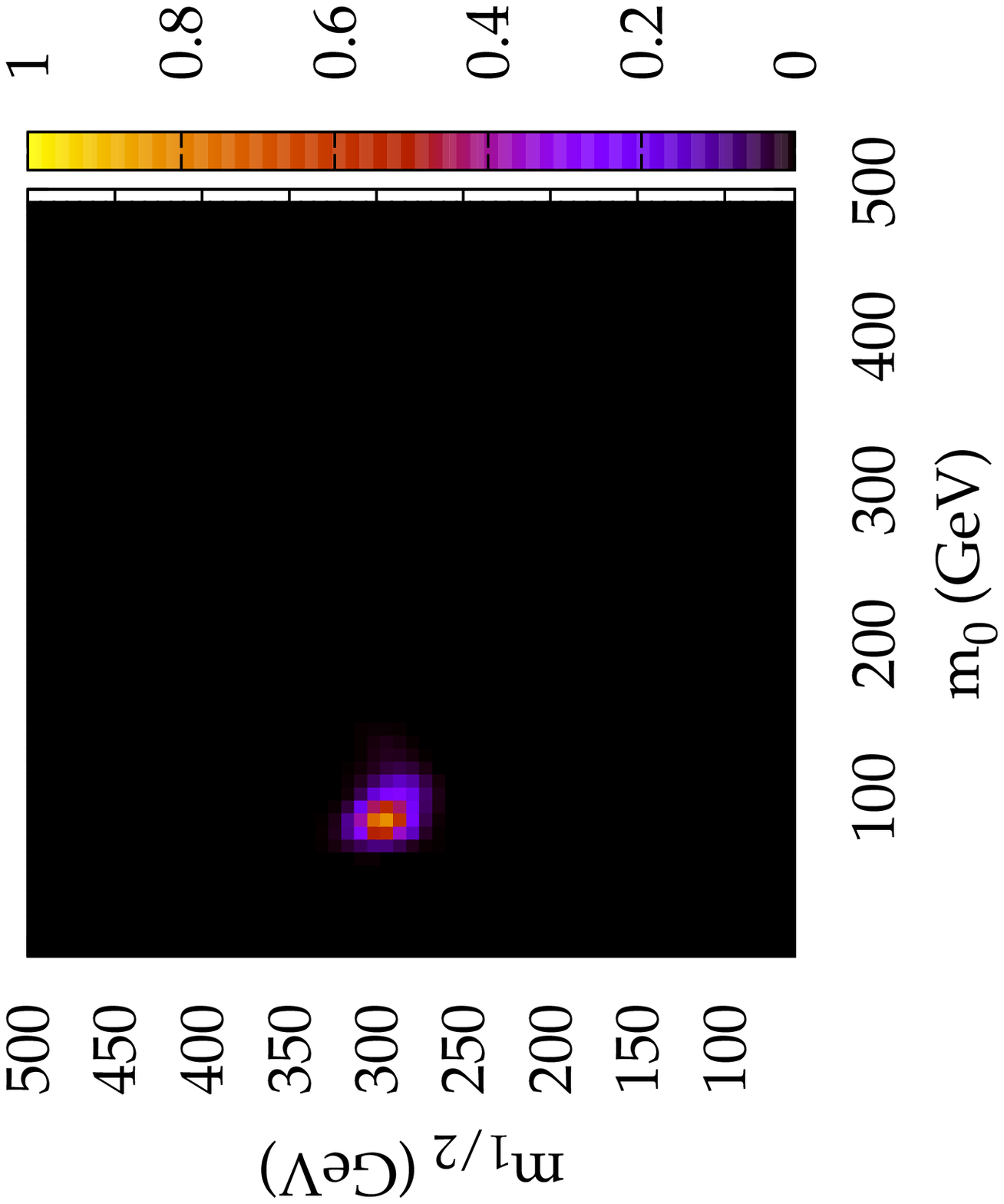}{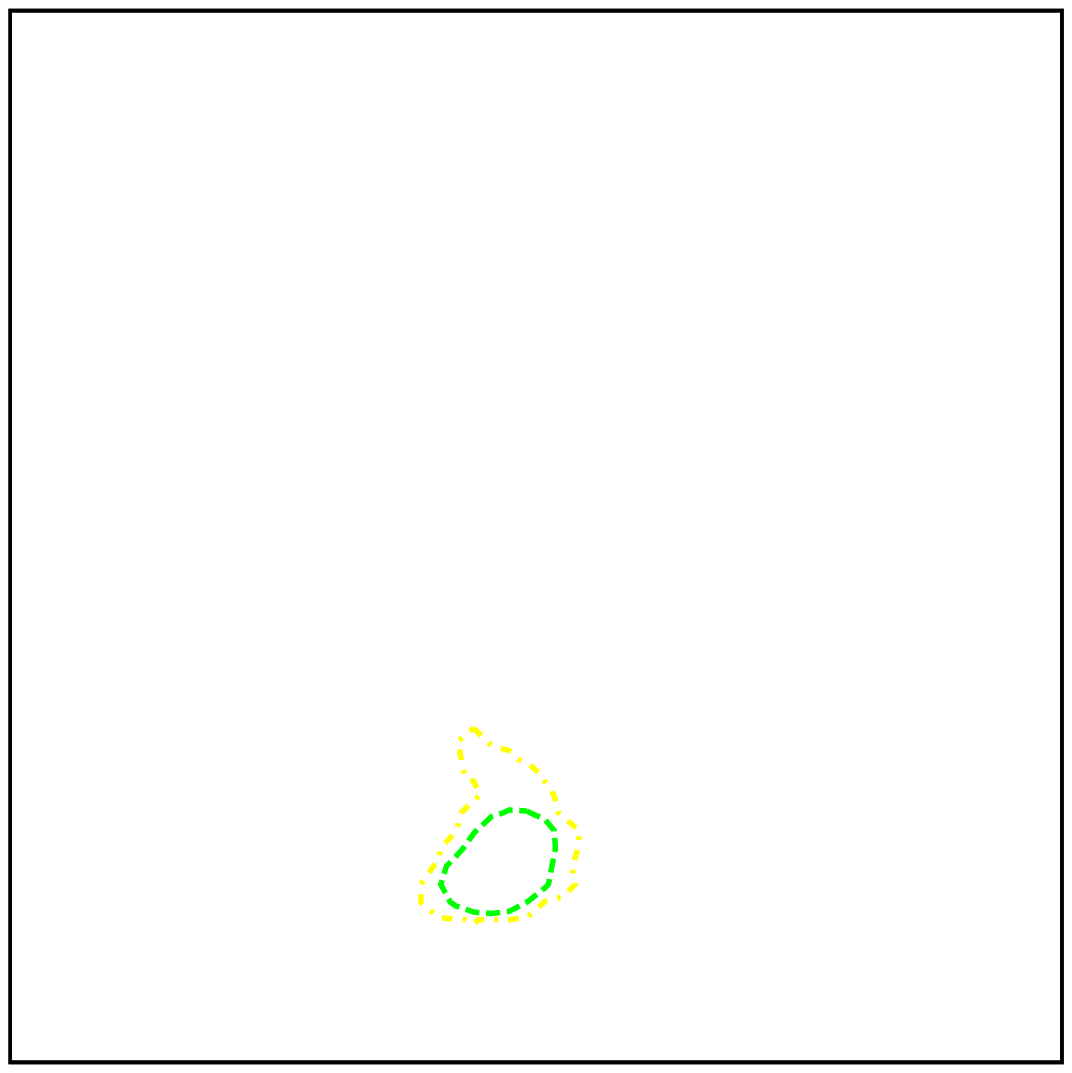}\caption{Posterior PDF for the CMSSM in the $m_0$-$m_{1/2}$ plane with log priors. The dashed green and dashed yellow contours show the 95\% Bayesian confidence intervals for log and flat priors respectively.\label{fig:cmssm2d}}
\end{center}
\end{figure}

We now discuss some features of the posterior distributions for the models
that are difficult to discriminate: the CMSSM and
mGMSB\@. We do not present the frequentist bounds upon the parameters using
$\Delta \chi^2$ because it has poor coverage properties~\cite{Bridges:2010de}. 
Figure~\ref{fig:cmssm2d} shows the 2D posterior for log priors in the
$m_0$-$m_{1/2}$ plane for the correct hypothesis for SUSY breaking, the
CMSSM\@. It 
also shows the 95\% Bayesian confidence interval contours for both sets of
priors. The posterior is a localised singled mode distribution, and the two
contours lie on top of one another, demonstrating prior independence in this
plane. This is not the case for the trilinear couplings $A_i$ which are not
well constrained by the edges, because these parameters have only a small
effect on the mass spectrum to which our fits are sensitive. Our posteriors
are in agreement
with previous fits of the CMSSM using kinematic
invariants~\cite{Lester:2005je,Roszkowski:2009ye,Fowlie:2011vf} 

\begin{figure}
    \begin{center}
\twographs{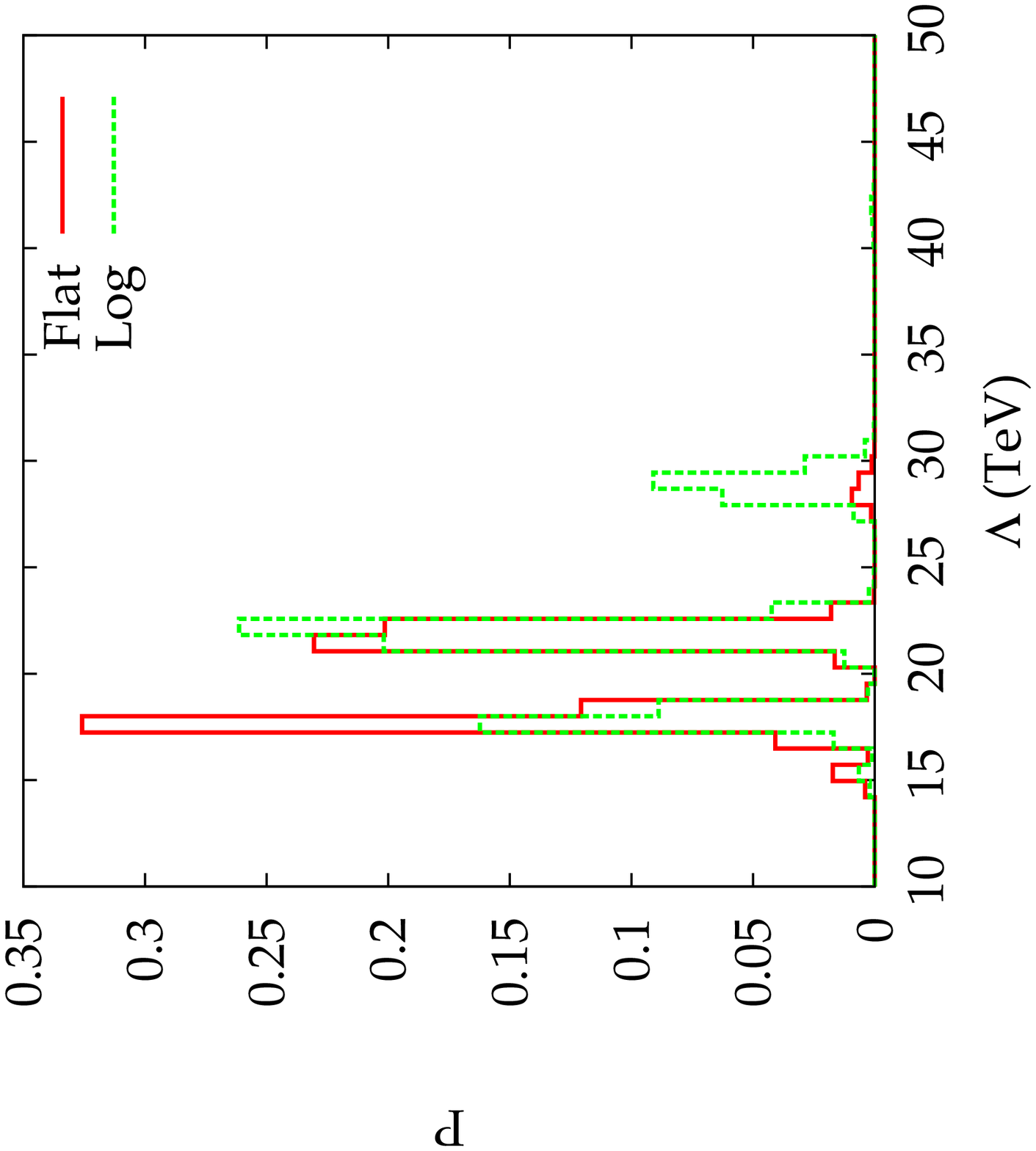}{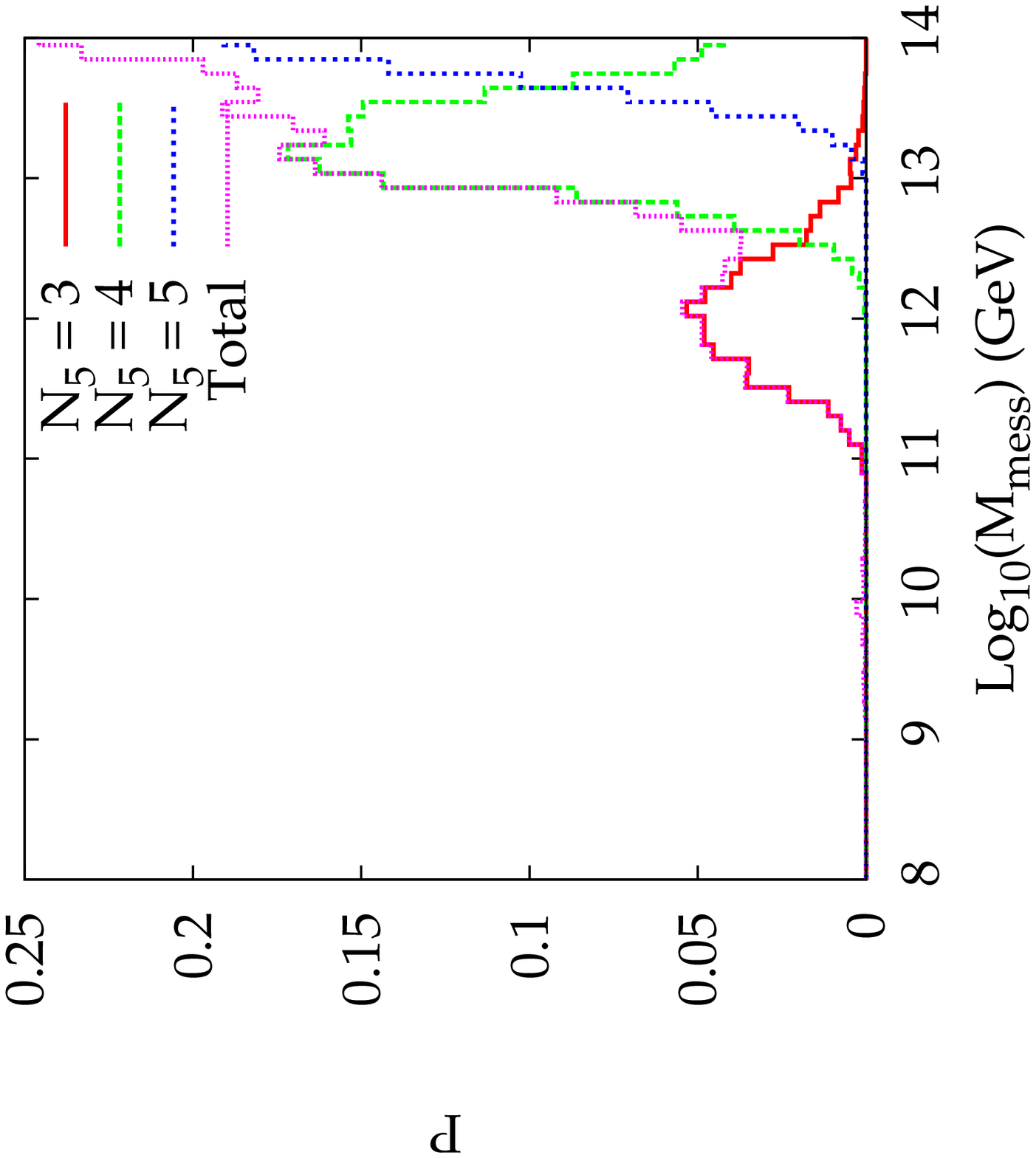}
      \end{center}
\caption{Posterior PDFs for mGMSB\@. The left-hand panel shows the 1D posterior
  for the mass scale $\Lambda$ for both priors and the right hand panel shows the 1D posterior for the  messenger mass $M_{mess}$ assuming logarithmic priors. The right-hand panel also shows the decomposition of the posterior according to messenger number $N_{mess}=3,4,5$.\label{fig:gmsb1d}} 
\end{figure}

Turning to mGMSB, Figure~\ref{fig:gmsb1d} shows 1D posteriors 
for the mass scale $\Lambda$ and the logarithm of the messenger
scale $\log_{10}(M_{mess})$ in GeV. 
We see from the left-hand panel that, in contrast to the CMSSM, the posterior
is strongly multi-modal, 
irrespective of prior. 
This is
because the physical masses in mGMSB are proportional to $N_{mess} \Lambda$,
and $N_{mess}$ is a discrete parameter. Each peak in the posterior for
$\Lambda$ corresponds to a different value for $N_{mess}$, with lower values
of $\Lambda$ being associated with higher values of $N_{mess}$, since their
product must be the mass scale given by the edge measurements. 
In the right-hand panel, we display the posterior of $M_{mess}$
separated according to
different values of $N_5$, as well as summed (`Total').
The $M_{mess}$ posterior extends down to 
$10^{11}$~GeV, having some substructure due to overlapping modes.
There is a positive correlation between $M_{mess}$ and
$N_{mess}$. From this we can infer that value of $N_{mess}$ larger than five
would only 
be favoured with unfeasibly high messenger scales. Low values of $M_{mess}$
require lower values of $N_{mess}$ in order to fit the data. Indeed, the {\sc
  MultiNest} algorithm identifies modes with $N_{mess}=1,2$, but these modes
are of poor fit quality compared with those of intermediate messenger number
$N_{mess}=3-6$. This is a salutary lesson that fitting a low dimensional model
to constraining data can still lead to a complicated mode structure in the
posterior.

\section{Conclusions \label{sec:con}}

We have evaluated the ability of the LHC, through the measurement of kinematic
end points in supersymmetric signals, to distinguish between different
models of supersymmetry breaking with a small number of parameters. We find
that the mAMSB and LVS models can be 
unambiguously 
discriminated from our CMSSM benchmark model by the end-points with just 1
fb$^{-1}$ of data. However, kinematic edges could not discriminate between
the best-fit CMSSM and mGMSB models, the spectra of which turn out to be
very similar (except for the gravitino mass, which is irrelevant for
LHC signals because the lightest neutralino is quasi-stable). 
Reassuringly, one reaches these conclusions whether or not one uses Bayesian or
frequentist statistics to perform the hypothesis test. 
This is additional confirmation
that the sparticle spectrum is sufficiently constrained by the measurements in
these models, and is confirmation of the fact that if a fit has sufficient
data, a Bayesian interpretation will be approximately prior independent and
give the  same results as a frequentist interpretation.
A previous study~\cite{Fowlie:2011vf}, found a significant prior
dependence in 
models of SUSY breaking that have more parameters than the CMSSM\@. This is not
so surprising given that the number of parameters would outnumber the number
of experimental constraints. In that case, we would not even be able to
calculate the $p-$value, since the number of degrees of freedom would be
negative, and the system is under constrained. 

The best-fit mGMSB and CMSSM spectra look remarkably similar, and a dedicated
analysis is required to see if LHC data can tell them apart, which looks {\em
a priori}\/ unlikely. 
It should be possible to use a future direct detection of
dark matter consistent with the CMSSM lightest neutralino mass in order to 
discriminate against mGMSB, whose gravitino LSP predicts zero direct detection
cross-section because it interacts too weakly\footnote{A recent study showed
that 
forthcoming ton scale direct detection experiments will probe the majority of
the CMSSM parameter space that currently fits indirect data
well~\cite{Bertone:2011nj}}.  
Another possible future extension of this work is to perform a simulated
experimental study of the best-fit mGMSB and CMSSM models, in order to see if
there are any observables that could discriminate between their rather close
spectra and decays. 
One could also attempt to answer the question: what sub-space of the CMSSM
parameter space predicts observables that are close to those of mGMSB? 
We would not expect mGMSB to be able to mimic a focus-point spectrum with
large $m_0$ but $m_{1/2}$ moderate for example, since this would result in a
rather hierarchical mass pattern, which the relatively compressed spectra of
mGMSB may find hard to reproduce. It is also true however, that the focus
point does not possess the golden decay chain and so different observables to
the ones studied here would have to be examined. 

Kinematic end-points of cascade decays are arguably the
best tool for discriminating different SUSY breaking models from LHC data,
since they are  
sensitive to the sparticle spectrum and do not require several hundred fb$^{-1}$
of integrated luminosity in order to parameterise the detector response well.
In the case that other cascades than the golden one assumed here are present
and identifiable, one would  
include their data. The fit is still likely to be dominated by the
constraints coming from the golden cascade, however. 
The golden cascade utilised here may not be present, even in the event of a SUSY
signal at the LHC. 
However, in that case other, less constraining cascades will be used but are
unlikely to provide the discriminating power that the golden one does. This
study is therefore an estimate of the maximum discriminatory power one could
have. 

In summary, although kinematic end-point data deliver important information on
the nature of SUSY breaking (discriminating against mAMSB, LVS and the CMSSM),
there still may exist degeneracies between some models (for example
mGMSB and the CMSSM SU3 benchmark point).
It would be interesting to see if a future linear collider
with a sufficient centre of mass energy could help separate the models. 
Refs.~\cite{Blair:2000gy,Blair:2002pg,Allanach:2004ud} demonstrate that using
LHC and linear collider data leads to accurate measurements of most of the
SUSY spectrum, as long as the relevant sparticles are kinematically accessible
at the linear 
collider. It was demonstrated how bottom-up renormalisation to high energies
allows checks on unification relations in different models. 
Given that top-down model discrimination is much more constrained than the
bottom-up analysis, and has many less
parameters, it seems plausible that model discrimination could be reached by
including the precise linear collider data. We leave a confirmation of this to
a future study.

\section*{Acknowledgements}
This work has been partially supported by STFC\@. We would also like to thank 
the IPPP for support under their associateship scheme. 

\bibliographystyle{JHEP}
\bibliography{wrongmodel}

\end{document}